\documentclass[prl,superscriptaddress,twocolumn]{revtex4-1}
\usepackage{amsmath}
\usepackage{dsfont}
\usepackage{bm}
\usepackage{graphicx}
\usepackage[applemac]{inputenc}
\usepackage[T1]{fontenc}
\usepackage{amssymb}
\usepackage{siunitx}
\usepackage{datetime}
\usepackage{color}
\usepackage{times}
\usepackage{lipsum} % for long equation
\usepackage{braket}
\usepackage[normalem]{ulem} % for strice-out
\usepackage{xr}
\usepackage{tikz}

\newcommand{\ii}{\mathrm{i}}
\newcommand{\ua}{\uparrow}
\newcommand{\da}{\downarrow}

\begin{document}
%%%%%%%%%%%%%%%%%%%%%%%%%%%%%%%%%%%%%%%%%%%%%%%%%%%%%%%%%%%%%%%%%%%%%%%%%%
\newcommand{\figdir}{.}
%\newcommand{\figdir}{figures}
%%%%%%%%%%%%%%%%%%%%%%%%%%%%%%%%%%%%%%%%%%%%%%%%%%%%%%%%%%%%%%%%%%%%%%%%%%
%%%affiliations
\newcommand{\freiburg}{Physikalisches Institut, Albert-Ludwigs-Universit\"{a}t-Freiburg, Hermann-Herder-Stra{\ss}e 3, D-79104, Freiburg, Germany}
\newcommand{\frias}{Freiburg Institute for Advanced Studies, Albert-Ludwigs-Universit\"{a}t-Freiburg, Albertstra{\ss}e 19, D-79104 Freiburg, Germany}
\newcommand{\TITLE}{Signatures of indistinguishability in bosonic many-body dynamics}

\title{\TITLE}
\author{Tobias Br\"unner}
\affiliation{\freiburg}
\author{Gabriel Dufour}
\affiliation{\freiburg}
\affiliation{\frias}
\author{Alberto Rodr\'iguez}
\affiliation{\freiburg}
\author{Andreas Buchleitner}
\email[]{a.buchleitner@physik.uni-freiburg.de}
\affiliation{\freiburg}

%\date{$Revision: 455 $, compiled \today, \currenttime}
%
\begin{abstract}

The dynamics of bosons in generic multimode systems, such as Bose-Hubbard models, is not only determined by interactions among the particles, but also by their mutual indistinguishability manifested in many-particle interference. We introduce a measure of indistinguishability for Fock states of bosons whose mutual distinguishability is controlled by an internal degree of freedom. We demonstrate how this measure emerges both in the non-interacting and interacting evolution of observables. In particular, we find an unambiguous relationship between our measure and the variance of single-particle observables in the non-interacting limit. A non-vanishing interaction leads to a hierarchy of interaction-induced interference processes, such that even the expectation value of single-particle observables is influenced by the degree of indistinguishability.

\end{abstract}
\maketitle
%%%%%%%%%%%%%%%%%%%%%%%%%%%%%%%%%%%%%%%%%%%%%%%%%%%%%%%%%%%%%%%%%%%%%%%%%%
%% main text
%%%%%%%%%%%%%%%%%%%%%%%%%%%%%%%%%%%%%%%%%%%%%%%%%%%%%%%%%%%%%%%%%%%%%%%%%%

Interference between indistinguishable particles is common to all many-particle quantum systems. 
Since the observation of the interference of two photons on a beamsplitter by Hong, Ou and Mandel (HOM) \cite{HOM:PRL87}, the highly non-trivial character \cite{MTichy:PRA11,YSRa:PRCLE13,SHTan:PRL13,NSpagnolo:Nat13,SAgne:PRL17,AJMenssen:PRL17} of many-particle interference has been demonstrated in extensive studies of photons transmitted through multimode beamsplitter arrangements \cite{MTichy:NJP12,SAaronson:ToC13,MTillmann:PRX15,VTamma:QIP16,VSShchesnovich:PRL16,JDUrbina:PRL16,MWalschaers:NJP16,ACrespi:NPho13,MABroome:Sci13,JBSpring:Sci13,Tillmann:NatComm13,JCarolan:NPho14,LLatmiral:NJP16,HWang:Nat17}. 
While these studies are restricted to non-interacting particles, it is clear that interference also occurs in the presence of interactions. This was shown for HOM-type interference \cite{andersson_quantum_1999,MKaufman:Sci14,WJMullin:PRA15,BGertjerenken:PL15,AKaufman:arXiv18}, in the dynamics of a bosonic Josephson junction \cite{MTichy:PRA12,GDufour:NJP17} or in quantum walks \cite{YLahini:PRA12,XQin:PRA14,Wang:PRA14,PPreiss:Science15,Wang:Ana16}. However, these results are limited to two particles or two external modes, and a systematic understanding of the interplay between interactions and many-particle interference in the time evolution of general many-particle systems is still lacking. This fundamental question is, however, key to a variety of complex quantum phenomena, such as dynamical equilibration after a quench \cite{MKaufman:Sci16,APolkovnikov:RMP11,JEisert:Nat15}, correlation formation \cite{Entanglement2,Entanglement1}, or transport in interacting many-body systems \cite{Wimberger,FMeinert:PRL14}. 
Furthermore, certification of the bosonic, fermionic, as well as (in)distinguishable character of particles \cite{CRoos:PRL17,MWalschaers:NJP16,MWalschaers:PRA16,Sciarrino,GiordaniNat18,VSShchesnovich:PRL16} could also be achieved by identifying the corresponding interference fingerprints in the (interacting) dynamics.

\begin{figure}
 \centering 
 \includegraphics[width=.8\columnwidth]{\figdir/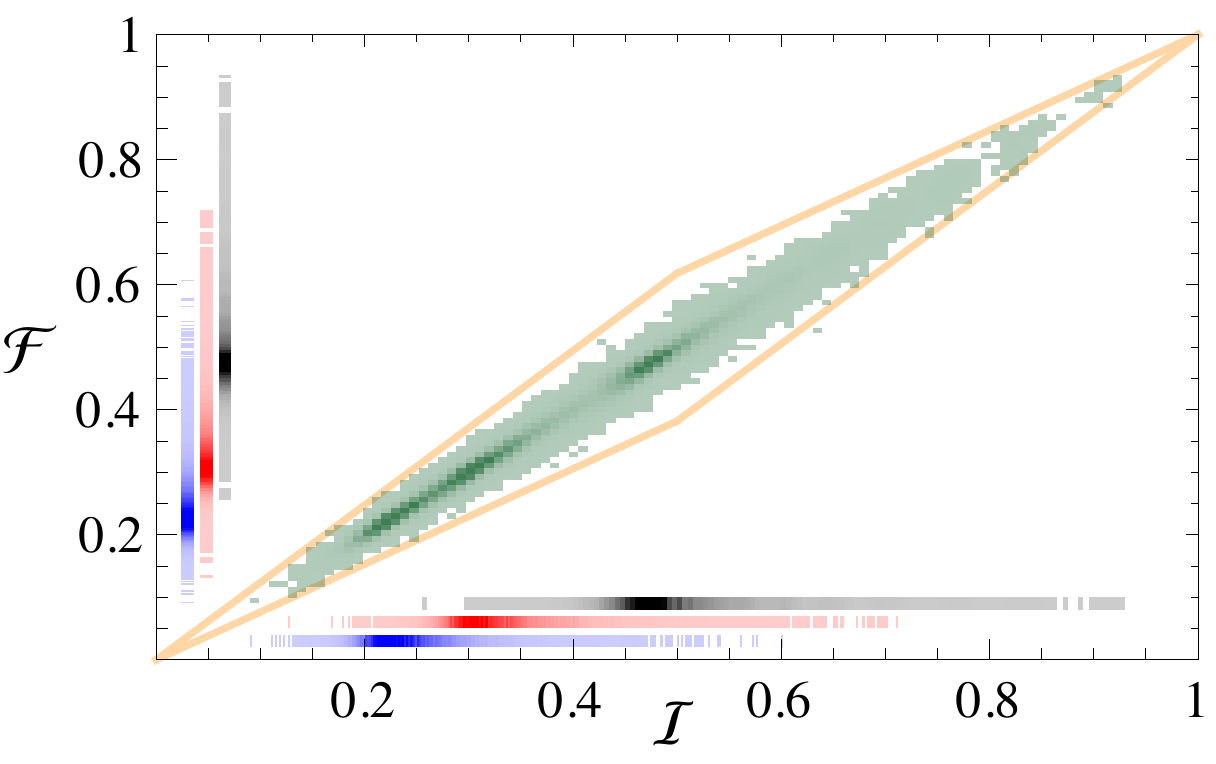}
 \caption{Density histogram of the normalized time-averaged variance $\mathcal{F}$ [Eq.~\eqref{eq:DefLoF}] of the on-site atomic density (at an arbitrary site) versus the DOI measure $\mathcal{I}$ [Eq.\eqref{eq:DefDoI}] of the initial Fock state in a non-interacting Bose-Hubbard system. We consider a total of $3\times10^5$ initial states sampled uniformly over the available Hilbert space of a system with $L=12$ sites and $N=24$ bosons of $S=2$ (black), $3$ (red), and $4$ (blue) distinct species. Projections of the histogram along the axes are shown independently for each $S$. Thick solid lines indicate our bound \eqref{eq:FIcorr} on the $\mathcal{F}$-$\mathcal{I}$ correlation.}
 \label{fig:FvsIL12N24}
\end{figure}

Hence, it is the purpose of this work to systematically explore the impact of particles' indistinguishability on the time evolution of interacting many-body systems. We consider bosons which occupy a discrete set of coupled modes and whose mutual (in)distinguishability is controlled by an additional internal degree of freedom.
First, we define a measure of the \emph{degree of indistinguishability} (DOI) of many-body configurations which is adapted to the study of interacting systems evolving continuously in time from an arbitrary initial Fock state. This is in contrast to other DOI measures introduced in non-interacting photonic scattering setups \cite{HGuise:PRA14,VShchesnovich:PRA14,VSShchesnovich1:PRA15,MTichy:PRA15,MWalschaers:PRA16}. We show that our measure has an intuitive interpretation in terms of two-particle interference. In the non-interacting case, it correlates directly with the \emph{variance} of experimentally accessible single-particle observables (1POs), as demonstrated in Fig.~\ref{fig:FvsIL12N24}. Remarkably, in the presence of interactions, the DOI is also imprinted on the bare \emph{expectation values} of 1POs.
     
%%%%%%%%%%%%%%%%%%
%%%%%%%%%%%%%%%%%%%%%%%

Let us consider a general many-particle system with a discrete set of mutually coupled external modes $l \in \{1, 2, \dots, L\}$ (e.g.~photonic input and output modes coupled via a beam splitter array, or tunnel-coupled sites in an optical lattice), and with a discrete set of internal states, or `species', $\sigma \in \{\sigma_1, \sigma_2, \dots, \sigma_S\}$ (e.g.~photon polarization 
or hyperfine states of atoms). For a many-body Fock state
$
  {\ket{\Psi} = \bigotimes_{l,\sigma} |N_{l,\sigma}\rangle},
$
with $N_{l,\sigma}$ bosons of species $\sigma$ in mode $l$, we propose the following quantitative measure of the DOI:
\begin{align}
	\mathcal{I} := \frac{ \sum_\sigma \sum_{m \neq n}N_{m,\sigma} N_{n,\sigma}}{\sum_{m \neq n} N_m N_n}\ .
	\label{eq:DefDoI}
\end{align}
Here, $N_l := \sum_\sigma N_{l,\sigma}$ denotes the total number of particles in mode $l$, such that $\mathcal{I} \in [0,1]$. This measure takes the value $1$ only when all particles are indistinguishable (i.e.~only one species is present). When each particle is of a different species (maximally distinguishable), then, consistently, $\mathcal{I}=0$. However, this minimum value is also reached when all particles of a given species occupy the same mode. According to our measure, the DOI does not solely depend on the repartition of particles among species, but also on how the species are distributed over the external modes \cite{IPR}. This interplay between external and internal degrees of freedom, although discussed for the indistinguishability of two photons \cite{MCTichy:FdP13,PTurner:arXiv16}, has not been clearly resolved in previously introduced DOI measures \cite{HGuise:PRA14,VShchesnovich:PRA14,VSShchesnovich1:PRA15,MTichy:PRA15,MWalschaers:PRA16}.

In the following, we demonstrate how this measure emerges in the dynamics of interacting and non-interacting systems. In order to assess the consequences of (in)distinguishability in the evolution, we require both the Hamiltonian and the measured observables to be \emph{species-blind}: they neither resolve, nor modify, the internal degree of freedom $\sigma$ of the particles \cite{SM}. In particular, the number of bosons per species is conserved.

\begin{figure}[t]
   \centering
   \includegraphics[width=.9\columnwidth]{\figdir/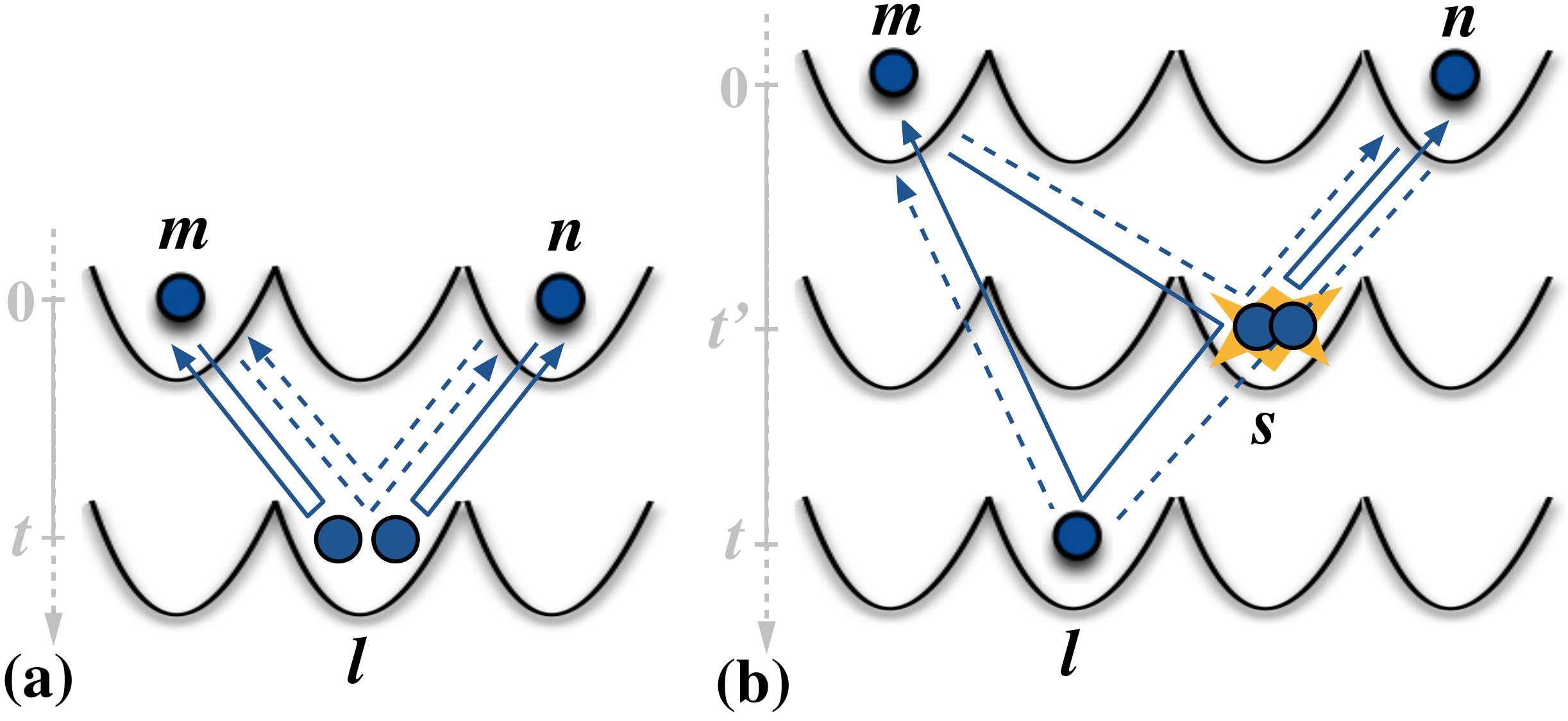}
   \caption{Two-particle paths of indistinguishable bosons: ladder (solid lines) and crossed (dashed lines). (a) Non-interacting:  
   	Processes ${(m,n)\rightleftarrows(l)}$ associated with the amplitude $C^{mn}_{llll}(t)$ in Eq.~\eqref{eq:LoF} contribute to the variance $\Delta \mathcal{N}_l(t)$ of the total density operator $\mathcal{N}_l$ of the \mbox{$l$-th} mode. 
   	(b) Interacting: Processes with amplitude $D_{ll}^{mn}(t)$  in Eq.~\eqref{eq:EV1POint} (which accounts for the interaction on all modes $s$, at times $t'\leqslant t$, before one of the particles visits mode $l$) contribute to the expectation value  $\langle\mathcal{N}_l(t,U)\rangle$.} 
   \label{fig:TwoParticlePaths}
\end{figure}

We first consider the \textit{non-interacting case}, where the Hamiltonian takes the general form of a species-blind 1PO, $\mathcal{H}_0=\sum_{i,j,\sigma}J_{ij}{a}^\dagger_{i,\sigma}{a}_{j,\sigma}$, and the time-evolution of the bosonic operators is given by the matrix elements $c_{lm}(t)$ of the single-particle unitary evolution operator: 
${a}_{l,\sigma}(t) = \sum_m c_{lm}(t) {a}_{m,\sigma}$.
Under these conditions, many-particle interference is known to manifest itself only on the level of two-particle or higher-order observables \cite{KMayer:PRA11}.
Indeed, the expectation value of a general species-blind two-particle observable (2PO), 
$\mathcal{O}_2=\sum_{i,j,k,l,\sigma,\rho} O_{ijkl} {a}_{i,\sigma}^\dagger{a}_{j,\rho}^\dagger{a}_{k,\sigma}{a}_{l,\rho}$, in a Fock state $\ket{\Psi}$ reads
\begin{align}
\braket{ \mathcal{O}_2(t) }_\Psi 
=& \sum_{i,j,k,l} O_{ijkl} \bigg[\notag \sum_{m,n} C^{mn}_{ijkl}(t) N_m (N_n-\delta_{mn})  \\
&+ \sum_{m \neq n, \sigma} C^{mn}_{jikl}(t) N_{m,\sigma} N_{n,\sigma}  \bigg].
\label{eq:ExpectTwoModeObs}
\end{align}
%%%%%%%%
The above expression can be interpreted as a sum over two-particle paths consisting of forward time evolution from the initial state, application of the observable and backward time evolution back to the same state \cite{SM}. The first line of Eq.~\eqref{eq:ExpectTwoModeObs} collects contributions of ``ladder'' paths, where the two particles initially in modes $m$ and $n$ return to their respective starting positions. These are associated with an amplitude $C^{mn}_{ijkl}(t):=  c_{im}^*(t) c_{jn}^*(t) c_{km}(t) c_{ln}(t)$ and a multiplicity $N_m N_n$. They are  common to all many-body configurations with the same initial total density distribution. Hence, they bear no information on the (in)distinguishability of the bosons. The second line in Eq.~\eqref{eq:ExpectTwoModeObs} represents additional ``crossed'' paths, where two particles \emph{of the same species $\sigma$}, initially in modes $m$ and $n$, are swapped, arriving in modes $n$ and $m$, respectively. Such processes have species- (i.e.~$\sigma$-) dependent multiplicities $N_{m,\sigma} N_{n,\sigma}$ and therefore bear information on the (in)distinguishability of the initially prepared many-particle configuration. Figure \ref{fig:TwoParticlePaths}(a) illustrates these ladder and crossed two-body processes for an observable which is local in the mode index. 

%%%%%%%%%%%%%%%%%%%%%%%%%%%%%%%%
%%%%%%%%%%%%%%%%%%%%%%%%%%%%%%%%%

The multiplicities of the crossed and ladder paths appear respectively in the numerator and denominator of our DOI measure [Eq.~\eqref{eq:DefDoI}], which therefore weighs the relative importance of the two types of processes in the expectation value of any 2PO [Eq.~\eqref{eq:ExpectTwoModeObs}]. We find that our measure manifests itself most directly when the 2PO under consideration is the square of a species-blind 1PO, $\mathcal{O}_1=\sum_{i,j,\sigma}O_{ij}{a}_{i,\sigma}^\dagger{a}_{j,\sigma}$, as this ensures that the factors $N_{m,\sigma}N_{n,\sigma}$ appear dressed by real and positive coefficients in Eq.~\eqref{eq:ExpectTwoModeObs}. In particular, we consider the variance $\Delta \mathcal{O}_1(t) :=\langle \mathcal{O}_1^2(t)\rangle_\Psi-\langle\mathcal{O}_1(t)\rangle_\Psi^2$ of on-site density operators $\mathcal{N}_l$, 
\begin{align}
	\Delta \mathcal{N}_l(t)  
	&= \sum_{m \neq n, \sigma} 
	 C^{mn}_{llll}(t) N_{m, \sigma} (N_{n, \sigma} + 1),
	\label{eq:LoF}
\end{align}
with amplitudes $C^{mn}_{llll}(t)=|c_{lm}(t)c_{ln}(t)|^2$. 
By averaging over time and subtracting the $\sigma$-independent contribution in Eq.~\eqref{eq:LoF}, the normalized time-averaged variance of the 1PO in the Fock state $\ket{\Psi}$ reads
\begin{align}\label{eq:DefLoF}
  \mathcal{F} &:= \frac{\overline{\Delta \mathcal{N}_l(t)} - \Delta_0}
  {\Delta_1 - \Delta_0}
  = \frac{\sum_{\sigma}\sum_{m \neq n} \overline{C^{mn}_{llll}(t)}\, N_{m,\sigma} N_{n,\sigma}}
  {\sum_{m \neq n}\overline{C^{mn}_{llll}(t)}\, N_m N_n}, 
\end{align}
where the overbar denotes time average, and $\Delta_{0,1}$ correspond to $\overline{\Delta \mathcal{N}_l(t)}$ in a state with the same total density distribution as $\ket{\Psi}$ but with $\mathcal{I}=0$ ($\Delta_{0}$) or $\mathcal{I}=1$ ($\Delta_{1}$)
(i.e.~in a fully distinguishable configuration, or in the state involving only one species, respectively \cite{ExpDOI}).
Comparison of Eqs.~\eqref{eq:DefDoI} and \eqref{eq:DefLoF} shows that, 
for a narrow distribution of the $\overline{C_{llll}^{mn}(t)}$  over $m\neq n$, 
the measurement of $\mathcal{F}$ directly gives access to the DOI. Specifically, we find that \cite{SM}
\begin{equation}
|\mathcal{F}-\mathcal{I}|\lesssim \frac{W_C}{\mu_C}\min(\mathcal{I},1-\mathcal{I}),
\label{eq:FIcorr}
\end{equation}
where $W_C$ and $\mu_C$ are, respectively, the standard deviation and the mean of the $\overline{C_{llll}^{mn}(t)}$ for all pairs $m\neq n$.

\begin{figure}
   \centering
  \includegraphics[width=.95\columnwidth]{\figdir/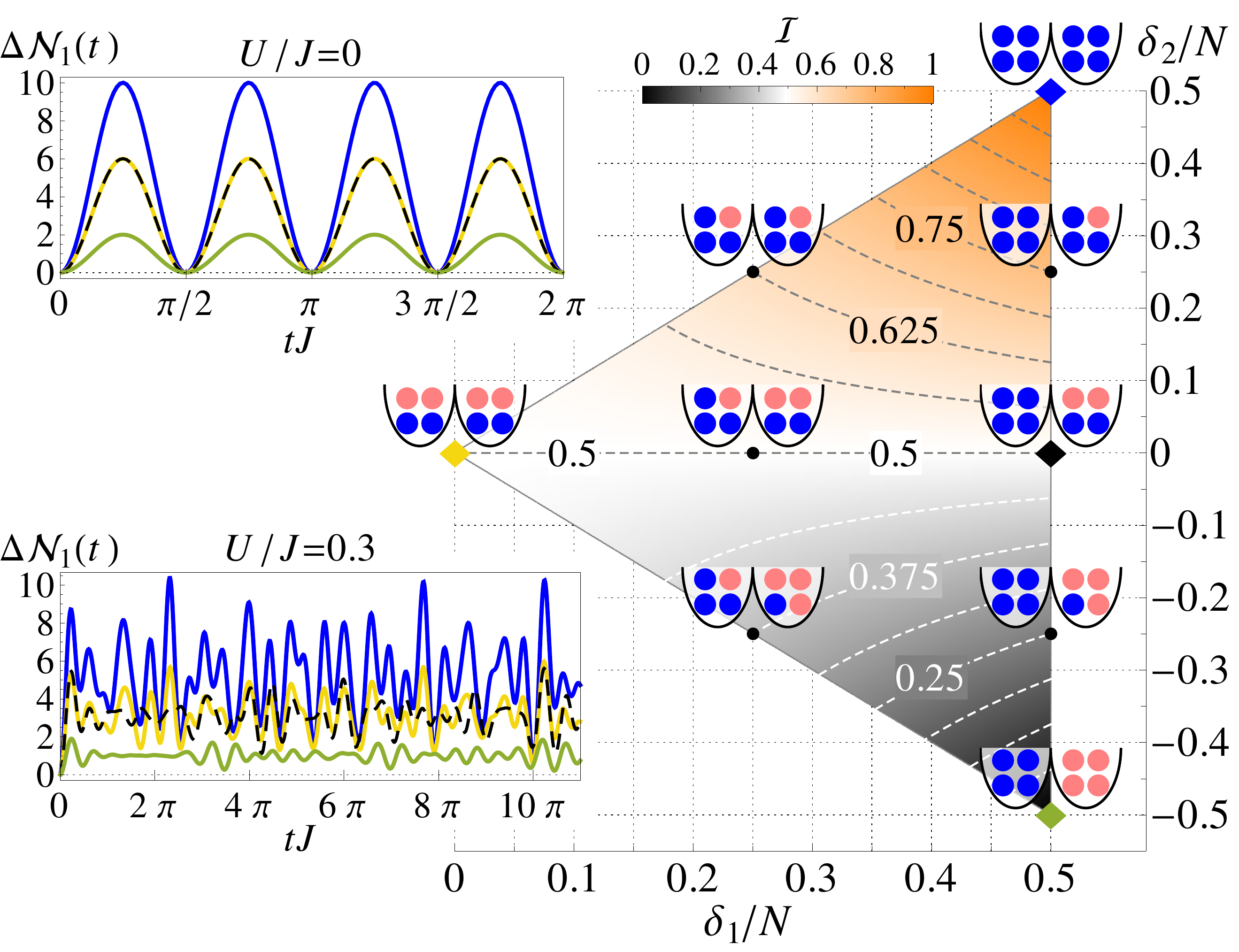}
  \caption{Density plot of the DOI for a two-species (blue/red) double well in the $\delta_1$-$\delta_2$ plane for $M=0$ [see Eq.~\eqref{eq:DoIL2}], including all nine non-equivalent configurations for $N=8$. Top and bottom insets show $\Delta \mathcal{N}_1(t)$ for four initial Fock configurations [$\mathcal{I}=1$ (blue, totally indistinguishable), $0.5$ (yellow, black dashed), and $0$ (green, maximally distinguishable)] for the non-interacting  and interacting ($U/J=0.3$) cases, respectively.}
   \label{fig:BJJ}
\end{figure}

It is instructive to study the behavior of our DOI measure in the special case of a two-mode system, such as a multi-component, species-blind, non-interacting Bose-Hubbard Hamiltonian (BHH) \cite{SM} with $L=2$ sites.
In any two-mode system, only one coefficient, $C_{llll}^{12}(t)$, contributes to $\mathcal{F}$, which therefore  reproduces \emph{exactly} the DOI measure $\mathcal{I}$. 
For two bosonic species $\sigma \in \{\uparrow, \downarrow\}$, and fixed total  particle number $N$, the configuration space of the system is determined by three parameters: the mode population imbalance, $M=N_1-N_2$, and the species imbalances per site, $\delta_1 = N_{1,\ua} - N_{1,\da}$ and $\delta_2 = N_{2,\ua} - N_{2,\da}$. The DOI measure then reads 
\begin{align}
\mathcal{I} = \frac{1}{2}+\frac{2 \delta_1 \delta_2}{N^2 - M^2}.
\label{eq:DoIL2}
\end{align}
For $M=0$, the space of non-equivalent Fock configurations is spanned by $\delta_1\in[0,N/2]$ and $|\delta_2|\leqslant \delta_1$, and is charted in Fig.~\ref{fig:BJJ} for $N=8$. According to Eq.~\eqref{eq:DoIL2}, having all particles of the same species [$\delta_1=\delta_2=N/2$] corresponds to $\mathcal{I}=1$, whereas complete spatial separation of the two species [$\delta_1=-\delta_2=N/2]$ implies $\mathcal{I}=0$. As shown in the top inset of Fig.~\ref{fig:BJJ}, these two initial states seed, respectively, maximum and minimum values of the density fluctuation $\Delta \mathcal{N}_1(t)$, as a direct consequence of the presence or absence of the two-particle crossed terms in Eq.~\eqref{eq:ExpectTwoModeObs} and Fig.~\ref{fig:TwoParticlePaths}(a).
Furthermore, all states with $\delta_2=0$ ---although having different species imbalance $\delta_1+\delta_2$--- have the same $\mathcal{I}=1/2$ and yield the same fluctuation of 1POs over time if the bosons do not interact. Conversely, states with equal species imbalance can exhibit different DOI values and hence dissimilar fluctuations.

Let us proceed to larger numbers of modes and species: We numerically demonstrate a remarkable $\mathcal{F}$-$\mathcal{I}$ correlation in a species-blind BHH with $L=12$ sites and a total of $N=24$ non-interacting bosons, as shown in Fig.~\ref{fig:FvsIL12N24}. We sample uniformly $10^5$ initial states out of the total available Fock space for each of the cases of $S=2$, $3$ and $4$ distinct species. For each state, $\mathcal{F}$ is calculated using Eq.~\eqref{eq:DefLoF}
and plotted versus the DOI value $\mathcal{I}$, together with
the bound provided by Eq.~\eqref{eq:FIcorr}. 
We observe that the $\mathcal{F}$-$\mathcal{I}$ correlation becomes even more pronounced for larger $L$ and/or $N$ \cite{DOIbound}. These results demonstrate that our DOI measure is at the core of the time-evolution of 2POs in non-interacting systems [see Eq.~\eqref{eq:ExpectTwoModeObs}], and furthermore, that it can be characterized from the \emph{variance}
of 1POs such as the on-site density of cold bosons in optical lattices.%.}

%%%%%%%%%%%%%%%%%%%%%%%%%%%%%%%%%%%%%%%%%%%%%%%%%%%%%%%%%%
%%%%%%%%%%%%%%%%%%%%%%%%%%%%%%%%%%%%%%%%%%%%%%%%%%%%%%%%%%%%%%

We now expand our analysis to the \textit{interacting case}, where, remarkably, the DOI  is revealed 
already in the \emph{expectation value} of 1POs.
To see this, we complement the Hamiltonian by a species-blind, two-body interaction term  $\mathcal{V}=\sum_{i,j,k,l,\sigma,\rho}V_{ijkl} {a}^\dagger_{i,\sigma}{a}^\dagger_{j,\rho}{a}_{k,\sigma}{a}_{l,\rho}$.
For simplicity, we elaborate on the case of contact `on-mode' interactions, $V_{ijkl}=(U/2)\delta_{ij}\delta_{jk}\delta_{kl}$; our subsequent results, however, are valid for the most general $\mathcal{V}$ \cite{SM}. In contrast to the non-interacting scenario, $a_{l,\sigma}(t)$ is now nonlinear in the initial creation and annihilation operators. Hence, in the Heisenberg picture, any 1PO develops, over time, a hierarchy of contributions in the form of two- and many-particle observables whose importance is weighted by the interaction strength. A perturbative treatment shows
%Solving the Heisenberg equation of motion perturbatively in $U$ shows that, over time
$\mathcal{O}_1(t,U) = \mathcal{O}_1(t,0)+(Ut) \mathcal{P}(t)+O((Ut)^2)$ \cite{SM}. 
Here, $\mathcal{O}_1(t,0)$ is a 1PO corresponding to the non-interacting evolution, with an expectation value  independent of the particles' (in)distinguishability. In contrast,
 $\mathcal{P}(t)$ is a 2PO with time-dependent matrix elements, and its expectation value reads
\begin{align}
\braket{ \mathcal{P}(t) }_\Psi 
=&\ 2 \Im \sum_{i,j} O_{ij} \bigg[\notag \sum_{m,n} D^{mn}_{ij}(t) N_m (N_n-\delta_{mn})  \\
&+ \sum_{m \neq n, \sigma} D^{nm}_{ij}(t) N_{m,\sigma} N_{n,\sigma}  \bigg],
\label{eq:EV1POint}
\end{align}
where $D^{mn}_{ij}(t)$ is the amplitude of the ladder and crossed two-particle paths arising due to the interaction \cite{SM}.
These are illustrated in Fig.~\ref{fig:TwoParticlePaths}(b) for the one-mode density $\mathcal{N}_l$ as 1PO.

By analogy with the result for 2POs in the non-interacting case [compare the structure of \eqref{eq:EV1POint} to that of \eqref{eq:ExpectTwoModeObs}], the DOI measure $\mathcal{I}$ can be identified in the expectation value of 1POs in the interacting case, dressed by the amplitudes $D^{mn}_{ij}(t)$. 
Interactions therefore imprint the DOI on the bare \emph{expectation value} of 1POs. This is demonstrated in Fig.~\ref{fig:IntEVn1}, where we show the expectation value $\langle\mathcal{N}_1(t,U)\rangle$ for the two-species two-site BHH \cite{BJJBE1,BJJBE2,BJJBE3,BJJBE4,BJJBE5,BJJBE6,BJJBE7,MTichy:PRA12,BJJBE8,GDufour:NJP17} subject to a tilt (to ensure a non-vanishing $Ut$ correction \cite{Usym}):
%%%%%%%%%%%%%%%%%%
\begin{figure}
 \includegraphics[width=\columnwidth]{\figdir/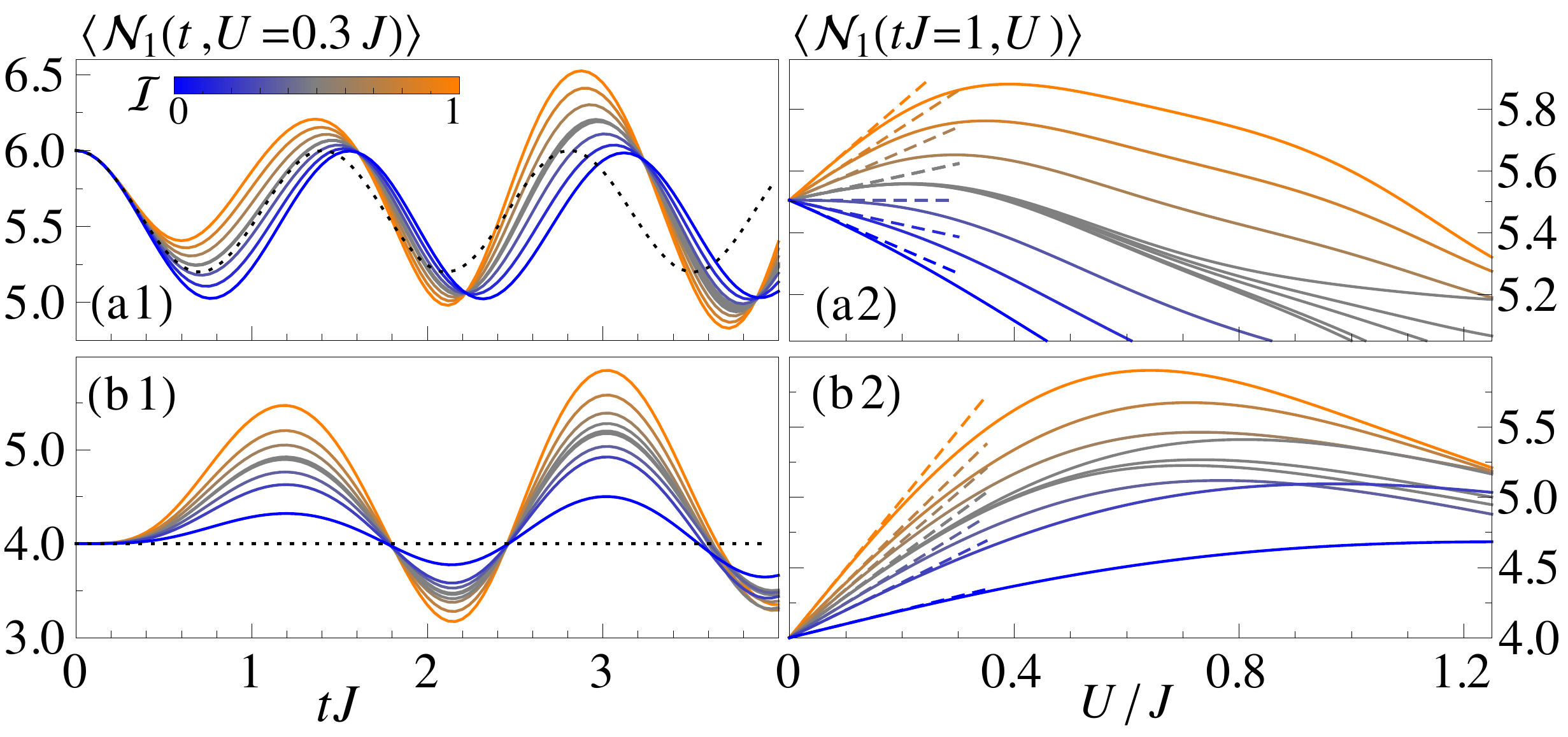}
 \caption{Expectation value $\langle \mathcal{N}_1(t,U)\rangle$ in the two-species double well with a tilt $F\mathcal{N}_2$, for $N=8$ and (a) $M=4, F=4J$, (b) $M=0, F=3J$, for all non-equivalent Fock states (color determined by DOI value $\mathcal{I}$ as indicated). The left (right) column shows the evolution versus $t$ ($U$) for fixed $U/J=0.3$ ($tJ=1$). Solid lines are numerical results, while dashed lines show the prediction of first order perturbation theory. Dotted lines in the left panels indicate the case $U=0$.} 
 \label{fig:IntEVn1}
\end{figure}
%%%%%%%%%%%%%%%%%%%
Within a regime of small $Ut$, which depends on the system under consideration, the evolution of the on-site density is well described by Eq.~\eqref{eq:EV1POint}. In particular, the initial slopes of the curves in Figs.~\ref{fig:IntEVn1}(a2) and (b2) are uniquely determined  by $\mathcal{I}$. For larger interaction strengths and/or times, higher-order terms contribute to the expectation value of the observable, which additionally probe three-particle and higher processes [causing, e.g., states with the same $\mathcal{I}=0.5$ to exhibit independent trajectories -- see panels (a2) and (b2) of Fig.~\ref{fig:IntEVn1}].  Nonetheless, the correlation between $\langle \mathcal{N}_1(t,U)\rangle$ and $\mathcal{I}$ persists beyond first order perturbation.
This suggests that our measure of the DOI based on two-particle paths remains meaningful even in the presence of higher-order processes.

\begin{figure}
 \includegraphics[width=\columnwidth]{\figdir/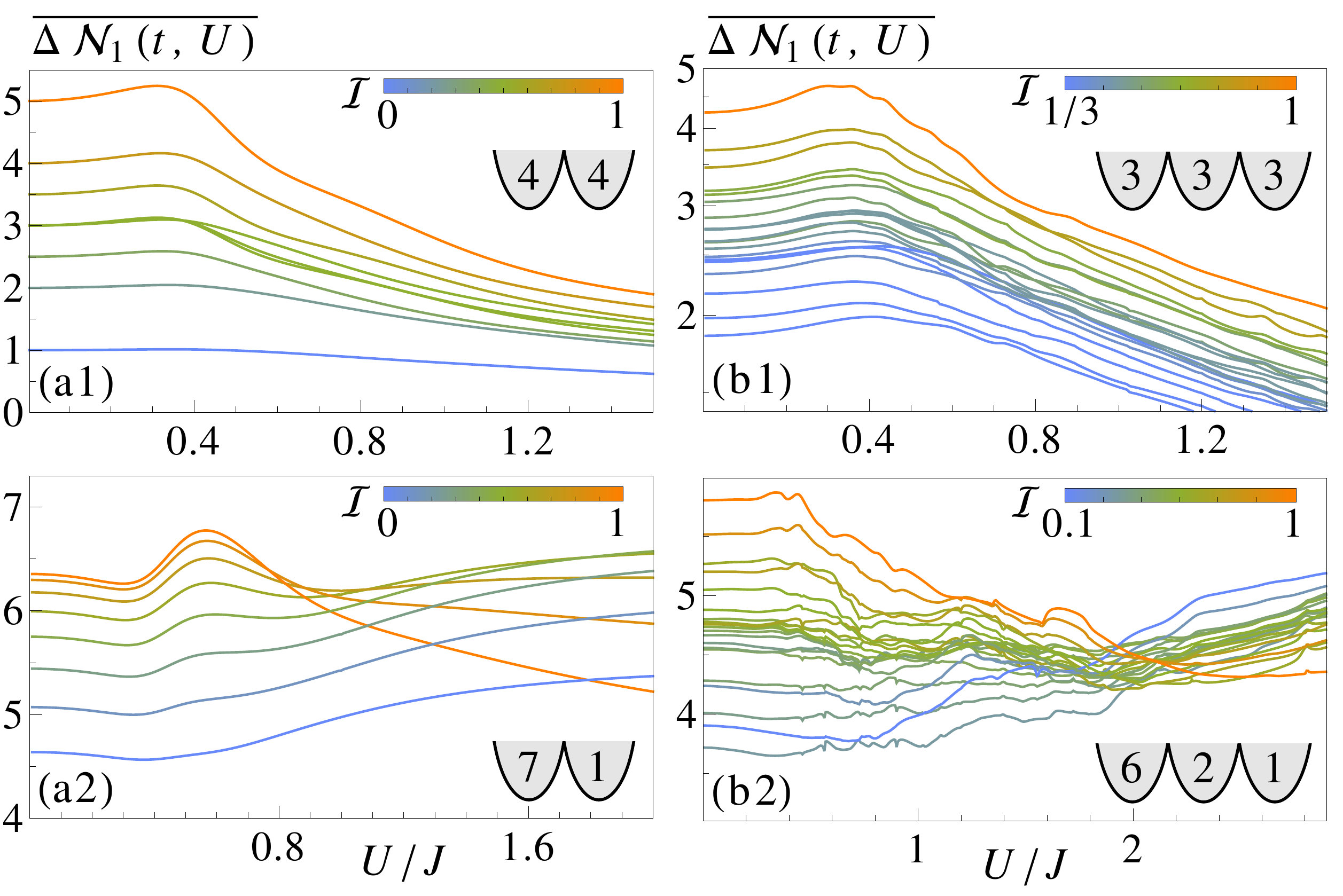}
 \caption{Time-averaged density fluctuation $\overline{\Delta\mathcal{N}_1(t,U)}$ versus interaction strength $U/J> 0$, for a BHH with two bosonic species: (a) $L=2, N=8$ 
 and (b) $L=3, N=9$, for the symbolically indicated initial densities and all non-equivalent Fock states (color determined by DOI value $\mathcal{I}$ as indicated). Note the vertical log scale in (b1). 
 }
 \label{fig:IntLOFn1}
\end{figure}

Indeed, also the long-time signals $\Delta\mathcal{N}_1(t,U\neq 0)$, although more involved than in the non-interacting case due to the appearance of extra frequencies (compare the top and bottom insets of Fig.~\ref{fig:BJJ}), indicate that the time-averaged density fluctuation still correlates with the DOI of the initial state.
This is demonstrated in Fig.~\ref{fig:IntLOFn1}, where we show $\overline{\Delta\mathcal{N}_1(t,U)}$ as a function of $U$ \cite{SM}. 
%%%%%%%%%%%%%%%%%%
%%%%%%%%%%%%%%%%%%%
For states with a homogeneous initial distribution of particles (first row of Fig.~\ref{fig:IntLOFn1}), one observes a striking correlation between $\overline{\Delta\mathcal{N}_1(t,U)}$ and $\mathcal{I}$ over the entire range of interaction strengths (also for $L>3$ -- not shown in the figure). For states with a strongly imbalanced initial distribution of particles (second row of Fig.~\ref{fig:IntLOFn1}), this correlation also holds for weak interactions, but is lost for larger values of $U$.  Closer inspection of the system's spectral structure shows that, in the regime of strong interactions, the dynamics is dominated by Fock states with the same interaction energy as the initial state, which, in the imbalanced case, include states with dissimilar density distributions [e.g. $\{7,1\}$ and $\{1,7\}$ in the double well]. The interaction-mediated higher-order processes connecting these states then contribute predominantly to $\overline{\Delta\mathcal{N}_1(t,U)}$, breaking the correlation to the DOI measure $\mathcal{I}$.
A detailed characterization of this effect will be the subject of future work.

We conclude by generalizing our DOI measure to superpositions of Fock states $\ket{\Psi}=\sum_j c_j\ket{\psi_j}$, where each term has the same total density distribution but a different number of particles per species.
The expectation value of a species-blind observable in such a state is additive, since by definition the observable cannot change the number of particles per species.
Thus, we can additively generalize our DOI measure as $\mathcal{I}_{\Psi}=\sum_j|c_j|^2\mathcal{I}_{\psi_j}$. 
For the exemplary Hong-Ou-Mandel state  $\ket{\Psi} = (\sqrt{\alpha}\,{a}^\dagger_{1,\uparrow} {a}^\dagger_{2,\uparrow}+ \sqrt{1-\alpha}\,{a}^\dagger_{1,\uparrow} {a}^\dagger_{2, \downarrow} ) \ket{\text{vac}}$, 
our measure $\mathcal{I}$ coincides with Mandel's indistinguishability parameter $\alpha$ \cite{LMandel:OL91}. 
Using the additivity property of $\mathcal{I}$,
 the effects of indistinguishability in various generalizations of the Hong-Ou-Mandel setup (e.g.~non-monotonicity in four-photon interference \cite{YSRa:PRCLE13,MTichy:PRA11}) are embedded in a more general framework.

We have introduced a measure of the degree of indistinguishability (DOI) of a many particle quantum state,
which is derived from the structure of two-particle transition 
amplitudes, and could be accessed in experiments by monitoring of the fluctuations
of one-particle observables. Our measure incorporates the significance of internal as well as
of external degrees of freedom for the DOI and for the associated 
many-particle interferences, and notably exploits the information encoded in the continuous dynamical 
many-particle evolution --- inaccessible in many-particle scattering scenarios. Our analysis also shows that 
{\it interaction-induced} interference reveals the DOI already in the {\it expectation
value} of single-particle observables, and that the DOI remains a meaningful concept in the presence 
of interactions.
The characteristic dynamical features observed here must have a structural counterpart in the underlying energy spectra and many-particle eigenstates, which deserve further investigation. We emphasize that our formalism and conclusions apply to general many-particle scenarios beyond the Bose-Hubbard model chosen to illustrate our results numerically.

%%%%%%%%%%%%%%%%%%%%%%
%acknowledgements
T.B. expresses gratitude to the German Research Foundation (IRTG 2079) for financial support and thanks Mattia Walschaers and Florian Meinert for helpful discussions. G.D. and A.B. acknowledge support by the EU Collaborative project QuProCS (Grant Agreement No. 641277). Furthermore, G.D. is thankful to the Alexander von Humboldt foundation. 
The authors acknowledge support by the state of Baden-W\"urttemberg through bwHPC and the German Research Foundation (DFG) through grant no INST 40/467-1 FUGG.

%%%%%%%%%%%%%%%%%%%%%%%%%%%%%%%%%%%%%%%%%%%%%%%%%%%%%%%%%%%%%%%%%%%%%%%%
%%%%%%%%%%%%%%%%%%%%%%%%%%%%%%%%%%%%%%%%%%%%%%%%%%%%%%%%%%%%%%%%%%%%%%%%
%  References
%%%%%%%%%%%%%%%%%%%%%%%%%%%%%%%%%%%%%%%%%%%%%%%%%%%%%%%%%%%%%%%%%%%%%%%%
\bibliographystyle{prsty}
%\bibliography{bibliography.bib}

\begin{thebibliography}{10}
\bibitem{IPR}
We note that $\mathcal{I}$ [Eq.~\eqref {eq:DefDoI}] can be written in terms of 
the inverse participation ratios (with respect to the external modes) of the 
density distributions, $\textrm {IPR}_\sigma
  =\sum _m N_{m,\sigma }^2/N_\sigma ^2$, where $N_\sigma $ is the total number
  of $\sigma $-particles, and $\textrm {IPR}=\sum _m N_m^2/N^2$. The DOI reads
  $\protect \mathcal {I} = \sum _{\sigma } \left ( N_\sigma ^2/N^2 \right )
  \left ( 1 - \textrm {IPR}_\sigma \right ) / \left ( 1 - \textrm {IPR}\right )
  $ and can thus be directly related to the degree of localization of each
  species over the external modes (occupation of one mode $\Rightarrow \protect
  \textrm {IPR}_\sigma =1$, homogeneous spreading $\Rightarrow \protect \textrm
  {IPR}_\sigma =L^{-1}$).
\bibitem{ExpDOI}
In an experimental estimation of $\protect \mathcal {F}$, the quantity $\Delta
  _0$ could be calculated theoretically if the parameters of the experimental
  Hamiltonian are known, without the need to engineer a state with $\protect
  \mathcal {I}=0$. The normalization factor, $\Delta _1-\Delta _0$, can be
  ignored if one compares ratios of $\protect \mathcal {F}$ to ratios of
  $\protect \mathcal {I}$.
\end{thebibliography}
%%%%%%%%%%%%%%%%%%%%%%%%%%%%%%%%%%%%%%%%%%%%%%%%%%%%%%%%%%%%%%%%%%%%%%%%  

\pagebreak
\clearpage

\begin{center}{\large{\bf Supplemental Material}}\end{center}

\section{Multi-species Bose-Hubbard model}
\label{sec:AppBHM}
As a particular realization of the class of systems discussed in the manuscript, we consider a one-dimensional Bose-Hubbard Hamiltonian (BHH) for many bosons which may be mutually distinguishable by an internal degree of freedom $\sigma$. The BHH describes bosonic atoms restricted to the first energy band of an optical lattice, and it contains a nearest-neighbor hopping term with typical energy $J$ and a two-body on-site interaction of strength $U$. In this case, the different species $\sigma$ may correspond to different hyperfine atomic states. 
The Hamiltonian of the system is chosen to be \emph{species-blind}, i.e.~it preserves the species type $\sigma$ and acts on all bosons in the same way: All bosons have the same hopping energy $J$ independently of $\sigma$, as well as the same inter and intra-species interaction, $U$. The measurement of a species-blind observable does not require to resolve the internal degree of freedom $\sigma$ of the bosons in the measurement process. The species-blindness condition is also known in the literature as isospecificity \cite{MTichy:PRA12}.
The total Hamiltonian reads $\mathcal{H}=\mathcal{H}_0+\mathcal{V}$, where 
\begin{align}
 \mathcal{H}_0 &= - J\sum_{\sigma,l} \left({a}_{l+1,\sigma}^\dagger {a}_{l,\sigma} + {a}_{l,\sigma}^\dagger {a}_{l+1,\sigma}\right),\label{eq:H0NN}\\
 \mathcal{V} &= \frac{U}{2}\sum_{l,\sigma,\rho} {a}^\dagger_{l,\sigma}{a}^\dagger_{l,\rho}{a}_{l,\sigma}{a}_{l,\rho},
\end{align}
in terms of bosonic creation (annihilation) operators ${a}_{l, \sigma}^\dagger$ (${a}_{l, \sigma}$), $\big[{a}_{l, \sigma}, {a}_{j, \rho}^\dagger\big] = \delta_{lj} \delta_{\sigma \rho}$, $\big[{a}_{l,\sigma}^\dagger, {a}_{j, \rho}^\dagger\big] = \big[{a}_{l,\sigma}, {a}_{j, \rho}\big] = 0$. 
We consider Hamiltonian $\mathcal{H}$ for a system comprising $L$ lattice sites in the presence of hard-wall boundary conditions.

%%%%%%%%%%%%%%%%%%%
\subsection{Dynamics in the non-interacting case}%
In the non-interacting case ($\mathcal{V}=0$), the dynamics of the system can be solved analytically. %[remember Eqs.~\eqref{eq:HeisenbergEqMotion} and \eqref{eq:SParticleSolution}]. We find for a system with $L$ sites and periodic boundary conditions (PBC)
For a generic Hamiltonian of the form $\mathcal{H}_0=\sum_{i,j,\sigma}J_{ij}{a}^\dagger_{i,\sigma}{a}_{j,\sigma}$ Heisenberg's equations of motion for the bosonic operators read (setting $\hbar\equiv1$)
\begin{align}
\frac{\mathrm{d}}{\mathrm{d}t} {a}_{l,\sigma}(t) = & i [\mathcal{H}_0,  {a}_{l,\sigma}(t)] \notag\\
= & -i \sum_m J_{lm} {a}_{m,\sigma}(t),
\end{align}
and therefore 
\begin{equation}\label{eq:clm}
{a}_{l,\sigma}(t)= \sum_m c_{lm}(t) {a}_{m,\sigma},
\end{equation}
where $c_{lm}(t)=\braket{ l|\exp(-i\mathcal{H}_0 t)|m}$ are the matrix elements of the single-particle evolution operator in the single-particle Wannier basis. 

For Hamiltonian \eqref{eq:H0NN} in the presence of hard-wall boundary conditions, one has 
\begin{align}
	c_{lm}(t) = 2\sum_{k=1}^L \sin\left(\frac{\pi k l}{L+1}\right)\sin\left(\frac{\pi k m}{L+1}\right) \frac{e^{2\ii tJ\cos\left(\frac{\pi k}{L+1}\right)}}{L+1}.
	\label{eq:BHSolOBC}
\end{align}
In order to asses the correlation between the indistinguishability measure $\mathcal{I}$ [Eq.~\eqref{eq:DefDoI}] and the level of fluctuation (LOF) $\mathcal{F}$ [Eq.~\eqref{eq:DefLoF}], we need to evaluate the time-averaged coefficients $\overline{C^{mn}_{llll}(t)}=\overline{|c_{lm}(t)c_{ln}(t)|^2}$.
The time average can be easily carried out analytically. However, the explicit evaluation of the resulting sums is rather involved. Nonetheless, 
the distribution of the values $\overline{C^{mn}_{llll}(t)}$ for $n\neq m$ and its dependence on the number of modes $L$ can be straightforwardly obtained numerically. 
%%%%%
\begin{figure}
 \includegraphics[width=.9\columnwidth]{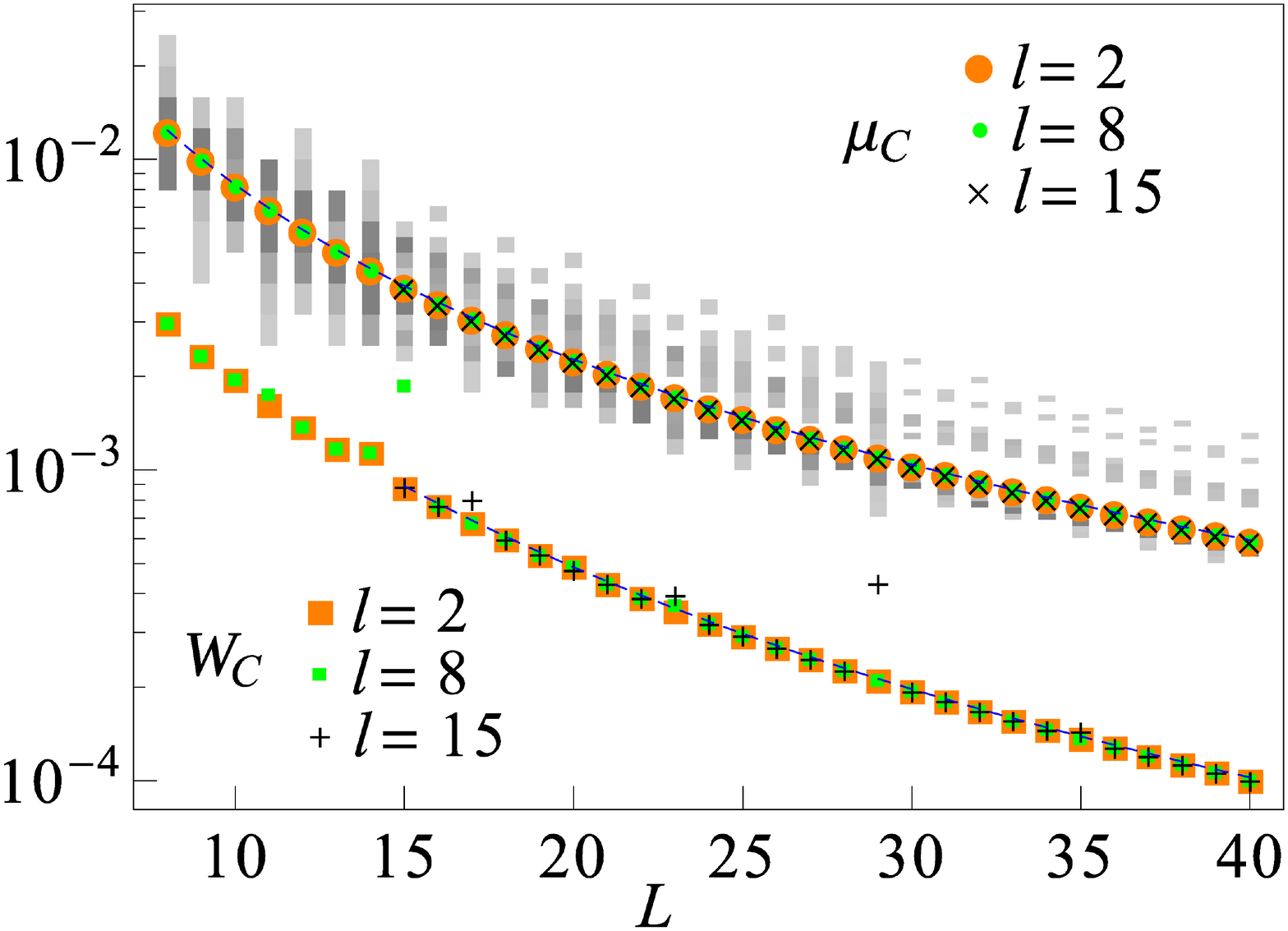}
 \caption{Scaling of the mean value $\mu_C$ and the standard deviation $W_C$ of the set of time-averaged coefficients $\overline{C^{mn}_{llll}(t)}$ for all pairs $m\neq n$ as a function of the number of sites $L$ (number of external modes). Top and bottom data show $\mu_C$ and $W_C$, respectively, for different sites $l=2, 8, 15$. The density plots in the background are the histograms of the coefficients for $l=2$. Dashed lines highlight the obtained fits given in Eqs.~\eqref{eq:fitmuc} and \eqref{eq:fitwc}.}
 \label{fig:C4coeff}
\end{figure}
%%%%%
This is shown in Fig.~\ref{fig:C4coeff}. The mean values $\mu_C$ are independent of the mode $l$ considered. The standard deviation $W_C$ also shows a common trend with $L$ independently of $l$. An exception occurs when the number of sites satisfy $L=2l-1$, i.e.~if $l$ is the centre of the mirror symmetry of the system, when we observe a jump in $W_C$ roughly by a factor of two. Nevertheless, this isolated resonant increase does not change the global decay of $W_C$ with $L$. 
As $L$ grows, the data is well fitted by the functions 
\begin{align}
 \mu_C &= 1.00 L^{-2} - 1.94 L^{-3} + 2.38 L^{-4} \label{eq:fitmuc}\\ 
   W_C &= 0.11 L^{-2} + 2.92 L^{-3} -22.50 L^{-4}, \label{eq:fitwc}
\end{align}
as demonstrated in Fig.~\ref{fig:C4coeff}. The ratio $W_C/\mu_c$ decreases with $L$ as
\begin{equation}
\frac{W_C}{\mu_C} = 0.11 + \frac{3.13}{L} - \frac{16.71}{L^{2}} - \frac{39.87}{L^{3}}+ O(L^{-4}),
\end{equation}
approaching the minimum value $0.11$ as $L\rightarrow\infty$.

%%%%%%%%
%%%%%%%%
\section{Estimation of the bounds for the $\boldsymbol{\mathcal{F}-\mathcal{I}}$ correlation}
\label{sec:AppError}

In order to derive a bound for the deviation of the normalized time-averaged variance $\mathcal{F}$ of the density operator $\mathcal{N}_l$ from the degree of indistinguishability (DOI) measure $\mathcal{I}$, we express both quantities as weighted averages $\{f\}$ over non-diagonal elements of matrices $f_{mn}$ defined on pairs of sites $(m,n)$:
\begin{align}
\{f\}&:=  \sum_{m \neq n}   \frac{N_m N_n}{\sum_{k \neq l}N_k N_l} f_{mn}.
\end{align}
The two quantities can be written as $\mathcal{I} = \{\eta\} $ and $\mathcal{F} =  \{\eta C\}/\{C\}$,
where
\begin{align}
\eta_{mn}&=  \frac{\sum_{\sigma}  N_{m, \sigma} N_{n, \sigma}}{N_m N_n},  &   &\quad  & C_{mn}&=  \overline{C^{mn}_{llll}(t)}
\end{align}
and the product $\eta C$ is performed entrywise: $(\eta C)_{mn}=\eta_{mn} C_{mn}$.
We now express the difference between $\mathcal{F}$ and $\mathcal{I}$ in two different ways:
\begin{align}
\mathcal{F}-\mathcal{I} =  \frac{\{\eta (C-\{C\})\}}{\{C\}}=\frac{\{(1-\eta) (\{ C\}-C)\}}{\{C\}}.
\end{align}
Given that $0\leqslant \eta_{mn}\leqslant 1$, we find, using successively both expressions of $\mathcal{F}-\mathcal{I}$:
\begin{align}
|\mathcal{F}-\mathcal{I}| &\leqslant   \frac{\{\eta |C-\{C\}|\}}{\{C\}} \approx \frac{ W_C}{\mu_C}\{\eta\},\\
|\mathcal{F}-\mathcal{I}| &\leqslant   \frac{\{(1-\eta) |C-\{C\}|\}}{\{C\}} \approx  \frac{ W_C}{\mu_C}(1-\{\eta\}),
\end{align}
where we have approximated the right hand side of the inequalities by assuming that $|C_{mn}-\{C\}|$ is of the order of the unweighted standard deviation $W_C$ of the distribution of $C_{mn}$, while the weighted average $\{C\}$ is approximated by its unweighted counterpart $\mu_C$. These approximations are valid for narrow enough distributions satisfying $W_C \ll \mu_C$.
The resulting estimation of the deviation of the normalized time-averaged variance from the DOI measure is thus:
\begin{align}
|\mathcal{F}-\mathcal{I}| \lesssim  \frac{ W_C}{\mu_C}\min(\mathcal{I},1-\mathcal{I}).
\label{eq:errorFvsI}
\end{align}
A rigorous bound can be obtained by noting that $|C_{mn}-\{C\}|\leqslant \max(C) -\min(C)$ and  	$\{C\}\geqslant \min(C)$ so that 
\begin{align}
|\mathcal{F}-\mathcal{I}| \leqslant  \frac{\max(C) -\min(C)}{\min(C)}\min(\mathcal{I},1-\mathcal{I}).
\end{align}

\section{Diagrammatic representation of time-dependent observables}

We give a diagrammatic interpretation of the time-dependent expectation values  \eqref{eq:ExpectTwoModeObs} and \eqref{eq:EV1POint}, both of which carry a signature of the DOI of the many-body configuration.
We recall the most general form of species-blind single-particle observables (1POs) and two-particle observables (2POs):
\begin{align}
\mathcal{A}_1&=\sum_{i,j,\sigma}A_{ij} {a}_{i,\sigma}^\dagger{a}_{j,\sigma},\label{eq:1PO}\\
\mathcal{A}_2&=\sum_{i,j,k,l,\sigma,\rho} A_{ijkl}{a}_{i,\sigma}^\dagger{a}_{j,\rho}^\dagger{a}_{k,\sigma}{a}_{l,\rho}, \label{eq:2PO} 
\end{align}
and their expectation values in the Fock state  $ {\ket{\Psi} = \bigotimes_{l,\sigma} |N_{l,\sigma}\rangle}$:
\begin{align}
\braket{\mathcal{A}_1}_\Psi &=\sum_{n}A_{nn} N_{n},\\
\braket{\mathcal{A}_2}_\Psi &=  \sum_{m,n} A_{mnmn}  N_m (N_n-\delta_{nm})\notag \\
&\qquad + \sum_{m \neq n, \sigma} A_{nmmn} N_{m,\sigma} N_{n,\sigma}.
\end{align}

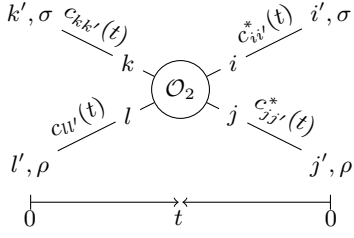
\begin{figure}
	\centering
	\begin{tikzpicture}
	\node[draw,circle] (O) at (0,0) {$\mathcal{O}_2$};
	\node (i) at (.7,.35) {$i$};
	\node (j) at (.7,-.35) {$j$};
	\node (k) at (-.7,.35) {$k$};
	\node (l) at (-.7,-.35) {$l$};
	\node (i') at (2,1) {$i',\sigma$};
	\node (j') at (2,-1) {$j',\rho$};
	\node (k') at (-2,1) {$k',\sigma$};
	\node (l') at (-2,-1) {$l',\rho$};
	
	\draw (O) -- (i);
	\draw (O) -- (j);
	\draw (O) -- (k);
	\draw (O) -- (l);
	
	\draw (i)  -- node[above,sloped]{$c_{ii'}^*(t)$} (i');
	\draw (j)  -- node[above,sloped]{$c_{jj'}^*(t)$} (j');
	\draw (k)  -- node[above,sloped]{$c_{kk'}(t)$} (k');
	\draw (l)  -- node[above,sloped]{$c_{ll'}(t)$} (l');
	
	\draw[|->] (-2,-1.5) node[below]{$0$}--  (-0.025,-1.5)node[below]{$t$};
	\draw[<-|] (0.025,-1.5)--(2,-1.5) node[below]{$0$};

	\end{tikzpicture}
	\caption{\label{fig:2POnonInt}Diagrammatic representation of a matrix element of $\mathcal{O}_2(t)$ in the absence of interactions. Two particle paths contributing to the expectation value \eqref{eq:ExpVal2PO} are obtained by connecting the left and right legs two-by-two. If $\sigma\neq \rho$, the only possibility is to join $i'$ with $k'$ and $j'$ with $l'$. If $\sigma= \rho$, one can also join $i'$ with $l'$  and $j'$ with $k'$.}
\end{figure}

We first consider a species blind 2PO, $\mathcal{O}_2$, of the form \eqref{eq:2PO}, with coefficients $O_{ijkl}$, evolving under a species-blind, non-interacting Hamiltonian $\mathcal{H}_0$.
%\begin{align}
%\mathcal{H}_0 = & \sum_{i,j,\sigma}J_{ij}{a}^\dagger_{i,\sigma}{a}_{j,\sigma},
%\end{align}	
In the Heisenberg picture, $\mathcal{O}_2(t)= \mathcal{U}_0^\dagger(t) \mathcal{O}_2  \mathcal{U}_0(t)$  is also a 2PO with matrix elements
\begin{align}
O_{i'j'k'l'}(t)=\sum_{i,j,k,l} O_{ijkl} c_{ii'}^*(t) c_{jj'}^*(t) c_{kk'}(t) c_{ll'}(t).
\end{align}
The coefficients $c_{lm}(t)$ are single-particle matrix elements of the evolution operator $\mathcal{U}_0(t)=e^{-i\mathcal{H}_0 t}$, as defined underneath Eq.~\eqref{eq:clm}.
This expression is represented graphically in Fig.~\ref{fig:2POnonInt}, where the legs on the left (right) of the diagram are associated with the forward (backward) time evolution $\mathcal{U}_0(t)$ ($\mathcal{U}_0^\dagger(t)$).
The corresponding expectation value reads
\begin{align}\label{eq:ExpVal2PO}
\braket{ \mathcal{O}_2(t) }_\Psi 
=& \sum_{i,j,k,l} O_{ijkl} \bigg[\notag \sum_{m,n} C^{mn}_{ijkl}(t) N_m (N_n-\delta_{mn})  \\
&+ \sum_{m \neq n, \sigma} C^{mn}_{jikl}(t) N_{m,\sigma} N_{n,\sigma}  \bigg].
\end{align}
Diagrammatically, it is obtained by identifying each leg on the left of the diagram to one on the right and to a populated mode in $\ket{\Psi}$.  Taking $i'=k'=m$ and $j'=l'=n$ leads to the ladder term, with coefficient $C^{mn}_{ijkl}(t)= c_{im}^*(t) c_{jn}^*(t) c_{km}(t) c_{ln}(t)$. For $\sigma=\rho$, one can also identify $j'=k'=m$ and $i'=l'=n$, yielding a crossed term with coefficient $C^{mn}_{jikl}(t)$.

\begin{figure}
	\centering
	\begin{tikzpicture}
	\node[draw,circle] (V) at (0,0) {$\mathcal{V}$};
	\node[draw,circle] (O) at (2.8,1) {$\mathcal{O}_1$};
	\node (i) at (.7,.35) {$i$};
	\node (j) at (.7,-.35) {$j$};
	\node (k) at (-.7,.35) {$k$};
	\node (l) at (-.7,-.35) {$l$};
	\node (p) at (2,1) {$p$};
	\node (o) at (3.6,1) {$o$};
	\node (i') at (5,1) {$i',\sigma$};
	\node (j') at (5,-1) {$j',\rho$};
	\node (k') at (-2,1) {$k',\sigma$};
	\node (l') at (-2,-1) {$l',\rho$};
	
	\draw (V)  -- (i);
	\draw (V) -- (j);
	\draw (V) -- (k);
	\draw (V) -- (l);
	\draw (O)  -- (p);
	\draw (O) -- (o);
	\draw (o) -- node[above,sloped]{$c_{oi'}^*(t)$}  (i');
	\draw (i)  -- node[above,sloped]{$c_{pi}(t-t')$} (p);
	\draw (j)  --  (2,-1)  -- node[above]{$c_{jj'}^*(t')$} (j');	
	\draw (k)  -- node[above,sloped]{$c_{kk'}(t')$} (k');
	\draw (l)  -- node[above,sloped]{$c_{ll'}(t')$} (l');

	\draw[|->] (-2,-1.75) node[below]{$0$}--  (2.775,-1.75)node[below]{$t$};
	\draw (0,-1.75)  node[]{|}  node[below]{$t'$} ;
	\draw[<-|] (2.825,-1.75)--(5,-1.75) node[below]{$0$};
	
	\end{tikzpicture}
	\caption{\label{fig:1POint} One of four diagrams contributing to $\mathcal{O}_1(t,U)$ to first order in the interaction. The single-particle observable is dressed by one interaction vertex, making it a two-particle observable.}
\end{figure}
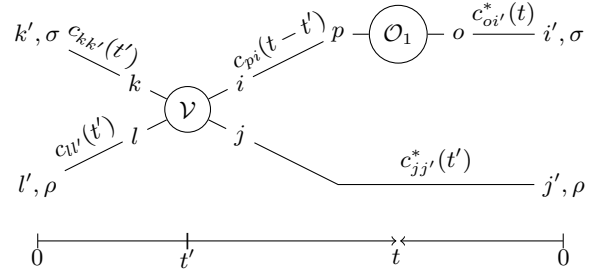

We now add an interaction term to the Hamiltonian, $\mathcal{H}= \mathcal{H}_0 +\mathcal{V}$,  in the form of a species-blind 2PO [see Eq.~\eqref{eq:2PO}], with coefficients $V_{ijkl}$ of order $U$.
To first order in $Ut$, the evolution operator $\mathcal{U}(t)=e^{-i\mathcal{H}t}$ can be written as
\begin{align}
\mathcal{U}(t)\approx& \ \mathcal{U}_0(t)-i \int_0^t  \mathcal{U}_0^\dagger(t-t') \mathcal{V}  \mathcal{U}_0(t')\ \mathrm{d}t'.
\end{align}		
Therefore, to the same order in $Ut$, a 1PO, $ \mathcal{O}_1$, of the form \eqref{eq:1PO}, with coefficients $O_{ij}$, evolves into
\begin{align}
\mathcal{O}_1(t,U)=&\ \mathcal{U}^\dagger(t) \mathcal{O}_1 \mathcal{U}(t) \approx\mathcal{O}_1(t,0)+(Ut) \mathcal{P}(t),\\
\intertext{with}
\mathcal{O}_1(t,0)= &\ \mathcal{U}_0^\dagger(t) \mathcal{O}_1  \mathcal{U}_0(t),\\
\mathcal{P}(t)=&\frac{i}{Ut} \left[ \int_0^t  \mathcal{U}_0^\dagger(t') \mathcal{V}  \mathcal{U}_0(t') \mathrm{d}t'  , \mathcal{O}_1(t,0)     \right].
\end{align}		
Here, $\mathcal{O}_1(t,0)$ is a 1PO and its expectation value is independent of the (in)distinguishability of the state. On the other hand, $\mathcal{P}(t)$ is a 2PO with matrix elements 
\begin{align}
P_{i'j'k'l'}(t)=&\frac{4}{Ut} \ \Im \sum_{ijklop}   V_{ijkl} O_{op} c_{oi'}^*(t)\notag \\
&\int_0^tc_{jj'}^*(t')c_{kk'}(t') c_{ll'}(t')c_{pi}(t-t')\ \mathrm{d}t'
\end{align}
(we have assumed, without loss of generality, that $V_{ijkl}$ is symmetric in the exchange of $i$ and $j$ and in the exchange of $k$ and $l$).
This matrix element is represented diagrammatically in Fig.~\ref{fig:1POint}.
For contact interactions $V_{ijkl}=\frac{U}{2}\delta_{ij}\delta_{jk}\delta_{kl}$, the corresponding  expectation value $\braket{\mathcal{P}(t)}_\Psi$ is given by
\begin{align}
\braket{ \mathcal{P}(t) }_\Psi 
=&\ 2 \Im \sum_{o,p} O_{op} \bigg[\notag \sum_{m,n} D^{mn}_{op}(t) N_m (N_n-\delta_{mn})  \\
&+ \sum_{m \neq n, \sigma} D^{nm}_{op}(t) N_{m,\sigma} N_{n,\sigma}  \bigg],
\end{align}
where
\begin{align}\label{eq:Dcoeff}
D^{mn}_{op}(t):= \frac{1}{t}\int_0^t dt' \sum_s |c_{sn}(t')|^2 c_{om}^*(t) c_{ps}(t-t') c_{sm}(t') .
\end{align}
This coefficient is associated with the two particle ladder process where [reading Eq.~\eqref{eq:Dcoeff} from right to left] one particle starts in mode $m$, moves to the interaction vertex $s$ in time $t'$ and  reaches mode $p$ at time $t$; it is then taken by the observable to mode $o$ before moving back to $m$. The other particle goes from mode $n$ to the interaction vertex $s$ and back.
The corresponding crossed term is obtained when the particles are exchanged at the interaction vertex.

%%%%%%%%%%%%
\section{Dynamics in systems with bipartite symmetry and on-site interactions}

Certain tight-binding Hamiltonians display a symmetry relating the dynamics for attractive and repulsive on-site interactions. Let $\mathcal{H}_U$ be a Hamiltonian with on-site interactions of strength $U$,
\begin{align}
\mathcal{H}_U= \sum_{i,j} J_{ij} a_{i}^\dagger a_{j}+ \frac{U}{2}\sum_{i}  a_{i}^\dagger a_{i}^\dagger a_{i} a_{i}.
\end{align}	
Note that we leave out the internal degree of freedom $\sigma$, which does not play a role in the following discussion.
Suppose that the single-particle modes can be split into two groups A and B, such that $J_{ij}=0$ if  $i$ and $j$ belong to the same group, including the case $i=j$ (bipartite symmetry). We define the unitary operator $\Pi$ by
\begin{align}
\Pi a_{i} \Pi^\dagger =\begin{cases}
a_i &\text{if } i\in \mathrm{A},\\
-a_i &\text{if } i\in \mathrm{B}.\\
\end{cases}
\end{align}	
The action of $\Pi$ on a Fock state $\ket{\Psi}$ is thus given by $\Pi\ket{\Psi}=\pm\ket{\Psi}$, where the sign depends on the parity of the number of particles on sites of type B. Therefore $\Pi$ is represented by a diagonal and real matrix in the Fock basis and  $\Pi=\Pi^\dagger=\Pi^{-1}$.
Since the kinetic term in the Hamiltonian only connects sites belonging to different groups, it changes sign under the action of $\Pi$. On the other hand, the interaction term is invariant. We therefore have $\Pi\mathcal{H}_U\Pi=-\mathcal{H}_{-U}$. The expectation value of an observable $\mathcal{O}$ in a Fock state $\ket{\Psi}$ thus obeys
\begin{align}
\braket{\Psi| e^{i \mathcal{H}_{U}t}   O e^{-i \mathcal{H}_{U}t}   |\Psi} &=\braket{\Psi|\Pi e^{-i \mathcal{H}_{-U}t}   \Pi  \mathcal{O} \Pi e^{i \mathcal{H}_{-U}t}  \Pi |\Psi}\notag \\
&=\braket{\Psi| e^{-i \mathcal{H}_{-U}t}   \Pi  \mathcal{O} \Pi e^{i \mathcal{H}_{-U}t}  |\Psi}. \label{eq:expectSym} 
\end{align}	
Given that the expectation value \eqref{eq:expectSym} is real, we can replace the right hand side of the equation by its complex conjugate. If the Hamiltonian is real in the Fock basis, we find that
\begin{align}
\braket{\Psi| e^{i \mathcal{H}_{U}t}   \mathcal{O} e^{-i \mathcal{H}_{U}t}   |\Psi} =\braket{\Psi| e^{i \mathcal{H}_{-U}t}   \Pi  \mathcal{O}^* \Pi e^{-i \mathcal{H}_{-U}t}  |\Psi}, 
\end{align}	
which relates the expectation value of $\mathcal{O}$ evolving under $\mathcal{H}_U$ to that of $\Pi  \mathcal{O}^* \Pi$ evolving under $\mathcal{H}_{-U}$.
If the observable satisfies $\Pi  \mathcal{O}^* \Pi=\mathcal{O}$, e.g.~for on-site density operators, its expectation value is invariant under switching the sign of interactions. In particular, all odd order terms vanish in the expansion of the expectation value in orders of the interaction strength.

\section{Time average of the variance}%

The time-averaged variance $\overline{\Delta \mathcal{O}(t)}$ of an arbitrary observable can be obtained in the Schr{\"o}dinger picture using the spectral decomposition of the state 
\begin{align}
\ket{\Psi(t)} = \sum_j c_j e^{-i E_j t} \ket{E_j},
\end{align}
where $E_j$ are eigenvalues of the Hamiltonian, $\ket{E_j}$ the corresponding eigenstates and 
$c_j = \braket{E_j|\Psi(0)}$ the weights of the initial state $\ket{\Psi(0)}$ in the eigenbasis. 
In the absence of energy degeneracies and gap degeneracies (no two pairs of eigenvalues are separated by the same energy gap), cancellation of all oscillating terms gives
\begin{align}\label{eq:time-avg-variance}
\overline{ \Delta \mathcal{O}(t) }=& \sum_j |c_j|^2 \bra{E_j}\mathcal{O}^2\ket{E_j}-\left( \sum_j |c_j|^2 \bra{E_j}\mathcal{O}\ket{E_j}   \right)^2\notag \\
&- \sum_{j\neq k} |c_j|^2 |c_k|^2 \ |\bra{E_j}\mathcal{O}\ket{E_k}|^2.
\end{align}
Energy or gap degeneracies yield extra terms in the expression of the time-averaged variance, which, in the Bose-Hubbard model, can lead to a discontinuity of $\overline{\Delta \mathcal{N}_l(U,t)}$ when $U$ goes to zero. For finite interactions $U>0$, we find that no such degeneracies contribute to $\overline{\Delta \mathcal{N}_l(U,t)}$ and the values given by the above formula agree with a direct numerical integration of the time signal $\Delta \mathcal{O}(t)$.
We emphasize that the irregular features observed in Fig.~\ref{fig:IntLOFn1} of the manuscript are genuine, and are obtained both using Eq.~\eqref{eq:time-avg-variance} and by numerical integration.

\end{document}